# Global Heuristic Search on Encrypted Data (GHSED)

Maisa Halloush, Mai Sharif

[1] Department of Computer Science, Al-Balqa Applied University,
Amman, Jordan
*Mhalloush@yahoo.com*

[2] Barwa Technologies,
Doha, Qatar
*Mai_shareef@hotmail.com*

**Abstract**
Important document are being kept encrypted in remote servers. In order to retrieve these encrypted data, efficient search methods needed to enable the retrieval of the document without knowing the content of the documents In this paper a technique called a global heuristic search on encrypted data **(GHSED)** technique will be described for search in an encrypted files using public key encryption stored on an untrusted server and retrieve the files that satisfy a certain search pattern without revealing any information about the original files. **GHSED** technique would satisfy the following: (1) Provably secure, the untrusted server cannot learn anything about the plaintext given only the cipher text. (2) Provide controlled searching, so that the untrusted server cannot search for a word without the user's authorization. (3) Support hidden queries, so that the user may ask the untrusted server to search for a secret word without revealing the word to the server. (4) Support query isolation, so the untrusted server learns nothing more than the search result about the plaintext.

***Key words:*** *Heuristic Table, Controlled Search, Query Isolation, hidden queries, false positive, hash chaining*

## 1. Introduction

With more and more files stored on not necessarily trusted external server, concerns about this file falling into the wrong hand grow (i.e. server administrator can read my file). Thus users often store their data encrypted to ensure confidentiality of data on remote servers, for more space, cost & convenience.

But what happen if the client wants to retrieve particular files (the files that satisfy certain search pattern or keyword)? A method is needed to search in the files for a particular keyword (search pattern) and only retrieve the files that contain that keyword. For example, consider a server that stores various files encrypted for "Alice" by others. A server wants to test whether the files contain the keyword "urgent" so that it could forward the file accordingly. "Alice", on the other hand does not wish to give the server the ability to decrypt all her files or even know anything about the search keyword.

A technique called a global heuristic search on encrypted data *(GHSED)* that enables server to search for a specific pattern on encrypted files without revealing any information to the untrusted server or any loss of data confidentiality will be defined and constructed, and its security would be proved. It also would be proved to have a minimal collision rate and stable construction time and it would also be proven to be applied to databases records, emails or audit logs.

*(GHSED)* technique is an enhancement over the *(HSED)* technique (Heuristic Search on Encrypted Data) technique [1], where they present "a new technique capable of handling large data keyword search in an encrypted document using public key encryption stored in untrusted server without revealing the content of the search and the document. The prototype provides a local search, minimizing communication overhead and computations on both the server and the client."[1].

*(HSED)* technique enables the server to efficiently search for a keyword without communication overhead since the message is encrypted and heuristic table construction is done on the client side. It also implies no additional computation overhead on the email server because no decryption is performed on the server since their model uses the public key cryptographic system. So unlike other techniques that use symmetric key cryptography, *(HSED)* technique reduces computation and communication overhead on the sever, in addition requires no additional computation except for simply calculating a hash function that serves as the address of an entry in the heuristic table."[1]

However, one of the disadvantages in *(HSED)* technique was that it deals with each document alone. When the sever search for a document that has a specific keyword it search all the document's heuristic table, this would be easy if the server has a small number of documents, but what if it has a large number of document? This would be too hard and needs a lot of time. Therefore, *(GHSED)* technique handles this problem by making a **GHT** (Global Heuristic Table) together with the **HT** (Heuristic Table) used in *(HSED)* technique.

Another disadvantage was that they drop the possibility of repeated word in a document; this would also be solved in *(GHSED)* technique, since this is a very common situation and has to be solved.

## 2. Global HEURISTIC SEARCH ON ENCRYPTED DATA *(GHSED)*

As shown in the previous section; *(HSED)* technique use heuristic table to make the search secure. Although the time needed to scan the heuristic table is encountered very little, it takes $O(M*e)$ to search in all documents in the server, cause it must go through one heuristic table per each document.

In this paper the same idea of *(HSED)* technique will be used; that is using public key cryptography, and search all keywords in document, but with one heuristic table for all documents in the server. This heuristic table will be named as Global Heuristic Table *(GHT)*, and it will contain information about each word exists in documents stored in the server.

Suppose "Alice" wants to store her encrypted documents in a form that could be searchable. To do this a Global Heuristic Table *(GHT)* will be used to contain every keyword in the documents stored in the server; each keyword will point to all documents in which the keyword located in. The pointers will be illusory pointers; that is the keyword will point to a binary array which contains every document number that the keyword exists in. It is clear that every document will be given a number before stored in the server.

When "Alice" wants to retrieve all documents which contain a specific keyword, she will send a trapdoor to the server. The server will use this trapdoor to search the Global Heuristic Table *(GHT)* and find the keyword, then retrieve the documents number which contain the keyword, and send the documents to "Alice".

In the next subsections we will illustrate how *(GHSED)* technique works, we will show the roll of every party involved in the search; that is the document generator, the searcher, and the server.

2.1 The Document Generator Side

When "Alice" wants to store an encrypted document in the server she will construct heuristic table **HT** which will be used by the server to embed it in the **GHT**. Then "Alice" will encrypt the document by "Alice's" public key $A_{pub}$. The document and the **HT** will be sent to the server. These steps are illustrated figure 2.1.

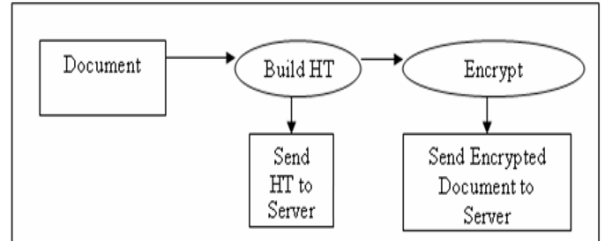

Fig 2.1: Steps of Constructing the Files.

*The heuristic table **HT** will be constructed as follows:*
For each keyword in the document a record in the heuristic table **HT** will be added as shown in Table 2.1

| [1] | KI($W_1$) | Ver-Key($W_1$) |
|---|---|---|
| [2] | KI($W_2$) | Ver-Key($W_2$) |
| [3] | KI($W_3$) | Ver-Key($W_3$) |
| . | . | . |
| . | . | . |
| . | . | . |
| Index$_i$ | KI($W_i$) | Ver-Key($W_i$) |

Table 2.1: Heuristic Table Header.

*Where:*
**Index$_i$ = H($W_i$)**   Calculated hash function used as index to both **HT** and **GHT** entries.

**KI($W_i$) =** $\sum_{j=1}^{n} ( chl_j(W_i) * chw_j( W_i ))$   The sum of each position of the $i^{th}$ character of the word multiplied by the character weight.

**EW$_i$**   Keyword $W_i$ Encrypted using $A_{pub}$

**Sum(EW$_i$) =** $\sum_{j=1}^{n} d_j$   The sum of digits of the encrypted keyword **EW$_i$**.

**Ver-key($W_i$) = KI($W_i$) || Sum(EW$_i$)**

### 2.2 The Server Side

Two operations will be done in the server side: the first one is storing the encrypted documents is a way to be searchable. The second one is to return documents which contain specific keyword when needed.

After building the **HT** which is related to the encrypted document, it will be sent to the server to be stored in it. The server will give a number to the document which will be used to define it. Then the server will embed the **HT** into **GHT**. The document with its **HT** will be stored in the server, as shown in figure 2.2.

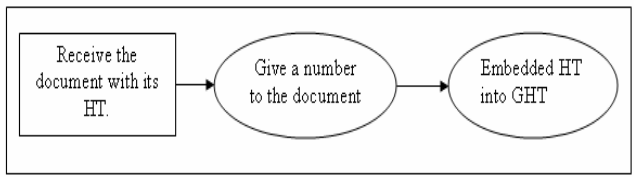

Fig 2.2: Steps done on the server side to store the documents.

**GHT** is a binary array contains information about each keyword exists in one or more documents stored in the server. If two words have the same index, the changing will be used. Each keyword in the **GHT** has a pointer to a binary array, which contains all documents numbers in which the keyword exists. Using the binary array will facilitate the operation done on the array.

Each entry in **HT** should be added to the **GHT**, such that, the index of both the **HT** and the **GHT** is the same. If no index exists in the **GHT** same as in **HT** then a new entry will be added in the **GHT**. This entry will contain both **KI** and **Ver-Key** and a pointer to a binary array which contains the document number.

If the index in **HT** exists in the **GHT**, then the entries of the table will be checked to see if we have the same word in the **GHT** by comparing both **KI** and **Ver-key**. If it is a new keyword then a new chain entry will be created and a pointer to a binary array which contains the document number. But if it is an existing keyword then just an entry containing the document number will be added in the binary array which contains the documents number.

### 2.3 Return Documents Which Contain a Specific Keyword

When "Alice" wants to retrieve the files that contain a specific keyword, she sends a trapdoor T $(T_{ew}, T_{KI})$ to the server

- $T_{ew}$: Keyword encrypted with "Alice's" public key

- $T_{KI} = \sum_{j=1}^{n} ( chl_j(W_i) * chw_j(W_i) )$

This trapdoor is used by the server to calculate an index in the global heuristic table **GHT**, that may the word locate
The position of the $i^{th}$ character of $W_i$ in the language multiplied by its position in the word.
Note that "Alice" then signs this trapdoor using her Private Key. Digital signature is used to allow the server identify that the trapdoor is sent by the recipient.
$Enc\ (Trapdoor(T_{ew}, T_{KI}), A_{priv})$

This will lead to an entry in the global heuristic table, the server can then retrieve the first and second column's entries from the heuristic table which is $< KI > < Ver\text{-}key >$ and calculate:

$Sum(T_{ew}) = \sum_{j=1}^{n} d_j$   The sum of digits of the first part of the trapdoor $T_w$

$Ver\text{-}key'(T_{ew}) = KI\ ||\ Sum(T_{ew})$

The calculated **Ver-key'($T_{ew}$)** will be compared to **Ver-key(W)**, which is the second table entry indexed by $H(T_{ew})$. If there is a match then the word exists in one of the files stored in the server, if not the index entry will be checked to see if there is a collided entries, if yes then the chain will be checked until a match will be found, if no match is found then the word does not exist.
If the word found in the **GHT** then the server will check the documents serial numbers that contains the specified word and send documents to "Alice". These steps are shown in figure 2.3.

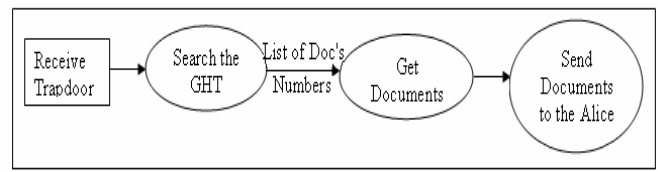

Fig 2.3.: Steps done on the server side for searching for the documents.

### 3. Results

**(GHSED)** technique algorithm mainly has two parts: embedding the heuristic table into the Global heuristic table (embedding operation) with take , and the searching part (search operation).

The embedding operation take $O(M*e)$ per each **HT**, where *M= maximum index number in* **HT**, *e= number of collided entries in the chain*. While the search operation will take $O(e)$ per each search, where *e= number of collided entries in the chain*.

It can be clearly noticed that there is overhead while embedding the **HT** into the Global one. This overhead will be increased by increasing the storing operations in the server, that is; if the need to store encrypted documents in the server done often, then there is an overhead on the server. But if the need to stored encrypted documents in the server is rarely done then the overhead may be ignored. On the other hand, the search time needed is very small in all cases as can be noticed.

The previous two parts of the *(GHSED)* technique algorithm were tested on a number of files that range in size from 10 KB to 5000 KB These files represent the heuristic table size in the first part of the algorithm, and represent the global heuristic table size which will be searched within it the second part of the algorithm.

The main interest is concentrated mainly on the time needed to embed the heuristic table into the global heuristic table, and on the search time needed to find in which documents a specific key word exists.

Figure 3.1 indicates that as the heuristic table size gets bigger, the time needed to embed it in the global heuristic table is higher, note that this process is being done on server.

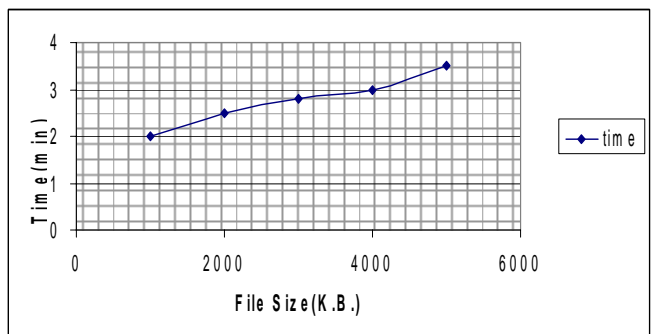

Fig.3.1: Time Needed to Embed Heuristic Table into the Global Heuristic Table.

Figure 3.2 indicates that no matter what the global heuristic table is, the search time will be constant.

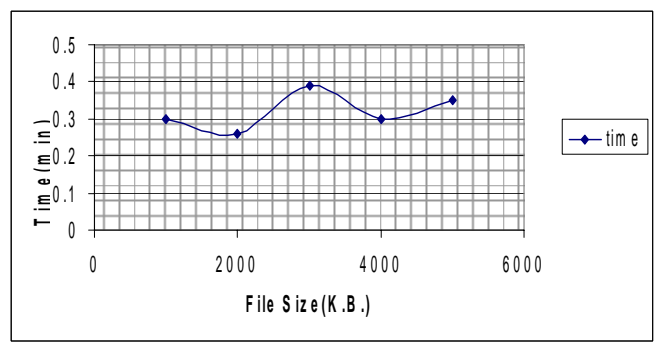

Fig.3.2: Time Needed To Search in the Global Heuristic Table.

## 4. Conclusion

*(GHSED)* technique enable search in the entire document for any keyword not just predefined keywords. It is efficient, fast and easy to implement. It minimizes communication and computation overhead. It can be applied to documents, emails, audit logs, and to database records. Any changes to the document can be detected because of the heuristic table. It can use hash chaining as it tightly links all entries in the array. It has no false positive; if the keyword appears to be in the document then it is in the document. It support hidden queries and query isolation. Finally, no one can detect the content of the document from the heuristic table so it is provably secure and it provides controlled searching.

But *(GHSED)* technique cannot be applied when the email server stores the emails compressed. Efficiency is dependent on the hash function to search for entries, so if the hash function is week, collision will occur more frequently and so the search will take longer.

Dealing with queries containing Boolean operations on multiple keywords remains the most significant and challenging open problem. Allowing general pattern matching, instead of keyword matching, also remains open.

**Maisa Halloush** Received the B.Sc. and M.Sc. scientific degrees in computer science 2002 and 2007, respectively. Her master thesis was about information security. She is continuing doing research in the same topic. Working currently as IT instructor in Al Quds College. She is also involved in writing books related to E-Business, Software Engineering, and Operating System.

**Mai Sharif** Received the B.Sc. and M.Sc. scientific degrees in computer science 1997 and 2005, respectively. She has more than 10 years experience in business analysis and software development. She has two published papers. Currently her areas of interest include information security, ripple effect in software modules, e-learning.

IJCSI